\documentclass[fleqn,12pt,twoside]{article}
\usepackage[headings]{espcrc1}
\readRCS
$Id: espcrc1.tex,v 1.2 2004/02/24 11:22:11 spepping Exp $
\usepackage{graphicx}
\usepackage[figuresright]{rotating}

\newcommand{\AmS}{{\protect\the\textfont2
  A\kern-.1667em\lower.5ex\hbox{M}\kern-.125emS}}

\title{New results from the PHOBOS experiment}

\runauthor{G. Roland for the PHOBOS collaboration}
\author{G. Roland for the PHOBOS collaboration\\
B.Alver$^4$,
B.B.Back$^1$,
M.D.Baker$^2$,
M.Ballintijn$^4$,
D.S.Barton$^2$,
R.R.Betts$^6$,
A.A.Bickley$^7$,
R.Bindel$^7$,
A.Budzanowski$^3$,
W.Busza$^4$,
A.Carroll$^2$,
Z.Chai$^2$,
V.Chetluru$^6$,
M.P.Decowski$^4$,
E.Garc\'{\i}a$^6$,
T.Gburek$^3$,
N.George$^2$,
K.Gulbrandsen$^4$,
S.Gushue$^2$,
C.Halliwell$^6$,
J.Hamblen$^8$,
G.A.Heintzelman$^2$,
C.Henderson$^4$,
I.Harnarine$^6$,
D.J.Hofman$^6$,
R.S.Hollis$^6$,
R.Ho\l y\'{n}ski$^3$,
B.Holzman$^2$,
A.Iordanova$^6$,
E.Johnson$^8$,
J.L.Kane$^4$,
N.Khan$^8$,
W.Kucewicz$^6$,
P.Kulinich$^4$,
C.M.Kuo$^5$,
W.Li$^4$,
W.T.Lin$^5$,
C.Loizides$^4$,
S.Manly$^8$,
A.C.Mignerey$^7$,
R.Nouicer$^{2,6}$,
A.Olszewski$^3$,
R.Pak$^2$,
I.C.Park$^8$,
C.Reed$^4$,
L.P.Remsberg$^2$,
M.Reuter$^6$,
E.Richardson$^7$,
C.Roland$^4$,
G.Roland$^4$,
L.Rosenberg$^4$,
J.Sagerer$^6$,
P.Sarin$^4$,
P.Sawicki$^3$,
I.Sedykh$^2$,
W.Skulski$^8$,
C.E.Smith$^6$,
M.A.Stankiewicz$^2$,
P.Steinberg$^2$,
G.S.F.Stephans$^4$,
A.Sukhanov$^2$,
A.Szostak$^2$,
J.-L.Tang$^5$,
M.B.Tonjes$^7$,
A.Trzupek$^3$,
C.Vale$^4$,
G.J.van~Nieuwenhuizen$^4$,
S.S.Vaurynovich$^4$,
R.Verdier$^4$,
G.I.Veres$^4$,
P.Walters$^8$,
E.Wenger$^4$,
D.Willhelm$^2$,
F.L.H.Wolfs$^8$,
B.Wosiek$^3$,
K.Wo\'{z}niak$^3$,
A.H.Wuosmaa$^1$,
S.Wyngaardt$^2$,
B.Wys\l ouch$^4$\\
$^1$~Argonne National Laboratory, Argonne, IL 60439-4843, USA\\
$^2$~Brookhaven National Laboratory, Upton, NY 11973-5000, USA\\
$^3$~Institute of Nuclear Physics PAN, Krak\'{o}w, Poland\\
$^4$~Massachusetts Institute of Technology, Cambridge, MA 02139-4307, 
USA\\
$^5$~National Central University, Chung-Li, Taiwan\\
$^6$~University of Illinois at Chicago, Chicago, IL 60607-7059, USA\\
$^7$~University of Maryland, College Park, MD 20742, USA\\
$^8$~University of Rochester, Rochester, NY 14627, USA\\
}

\begin{document}

\maketitle

\begin{abstract}
Over the past five years, PHOBOS has collected data on proton-proton, deuteron-gold, Au+Au and Cu+Cu 
collisions, covering a wide range of collision energy, collision centrality and system size. 
Using these data, we have identified scaling features that are common for all types of high-energy collisions,
as well as collective effects that are unique to the conditions created in collisions of relativistic nuclei.
In this paper, we will focus on recent results obtained for collisions of Cu nuclei. Both in terms of universal 
features of particle production, and in the development of truly collective effects,
the results for Cu nuclei confirm and extend our present understanding of nuclear collisions at the highest 
energies. In addition, we will describe recent unique results on multiplicity fluctuations and particle production
at very low transverse momenta.
\end{abstract}

\section{Overview}

PHOBOS is one of four experiments at the Relativistic Heavy Ion Collider (RHIC) at Brookhaven National Lab, studying collisions
of heavy nuclei at energies up to $\sqrt{s_{NN}} = 200$~GeV. Our goal is to study the properties of strongly interacting
matter at extreme temperature and density, where ab-initio numerical QCD calculations predict a phase transition to 
a system dominated by quark and gluon degrees of freedom. Our findings from the first four years of these studies
have been summarized in \cite{phobos_whitepaper}. There we argued that in Au+Au collisions, a dense, highly interacting system 
is created, with energy densities far in excess of the critical values for the QCD phase transition. 

\section{The PHOBOS experiment}

A detailed description of the PHOBOS detector can be found in \cite{phobos_nim}. The main components of the apparatus 
are a two-arm high-resolution magnetic spectrometer near mid-rapidity and a nearly 4-$\pi$ charged particle multiplicity detector
covering $| \eta | < 5.4$. Both the spectrometer and multiplicity array use analog readout silicon pad sensors for particle 
detection, with a total of 135000 channels. The silicon detectors are complemented by scintillator based trigger and 
time-of-flight counters, and forward calorimeters used for event characterization by detecting spectator nucleons.
In five runs at RHIC, PHOBOS has collected more than $10^9$ events to tape. An overview of the available datasets can be found in Table~\ref{table1}. A key component of the experiment
that is not usually discussed is the unique offline computing architecture that PHOBOS has established at the RHIC 
computing facility (RCF). We decided at an early stage  to implement a distributed disk solution, equipping all nodes in our RCF Linux farm
with cheap disk drives, for a total storage capacity of close to 100~TByte. That allows us to keep our entire physics dataset
available on disk for analysis at all times. In addition, we developed effective data management software and collaborated in the 
development of the PROOF extensions of ROOT \cite{root}, allowing us to use the main RCF cluster for interactive, parallel analysis of 
the large Cu+Cu and Au+Au data samples. The effective bandwidth achieved in these analyses using PROOF at RCF exceeds 5~GByte/sec.
Details of the PHOBOS computing architecture can be found in \cite{proof} and a forthcoming publication.

\begin{figure}[t]
\begin{minipage}[t]{80mm}
\includegraphics[width=7cm]{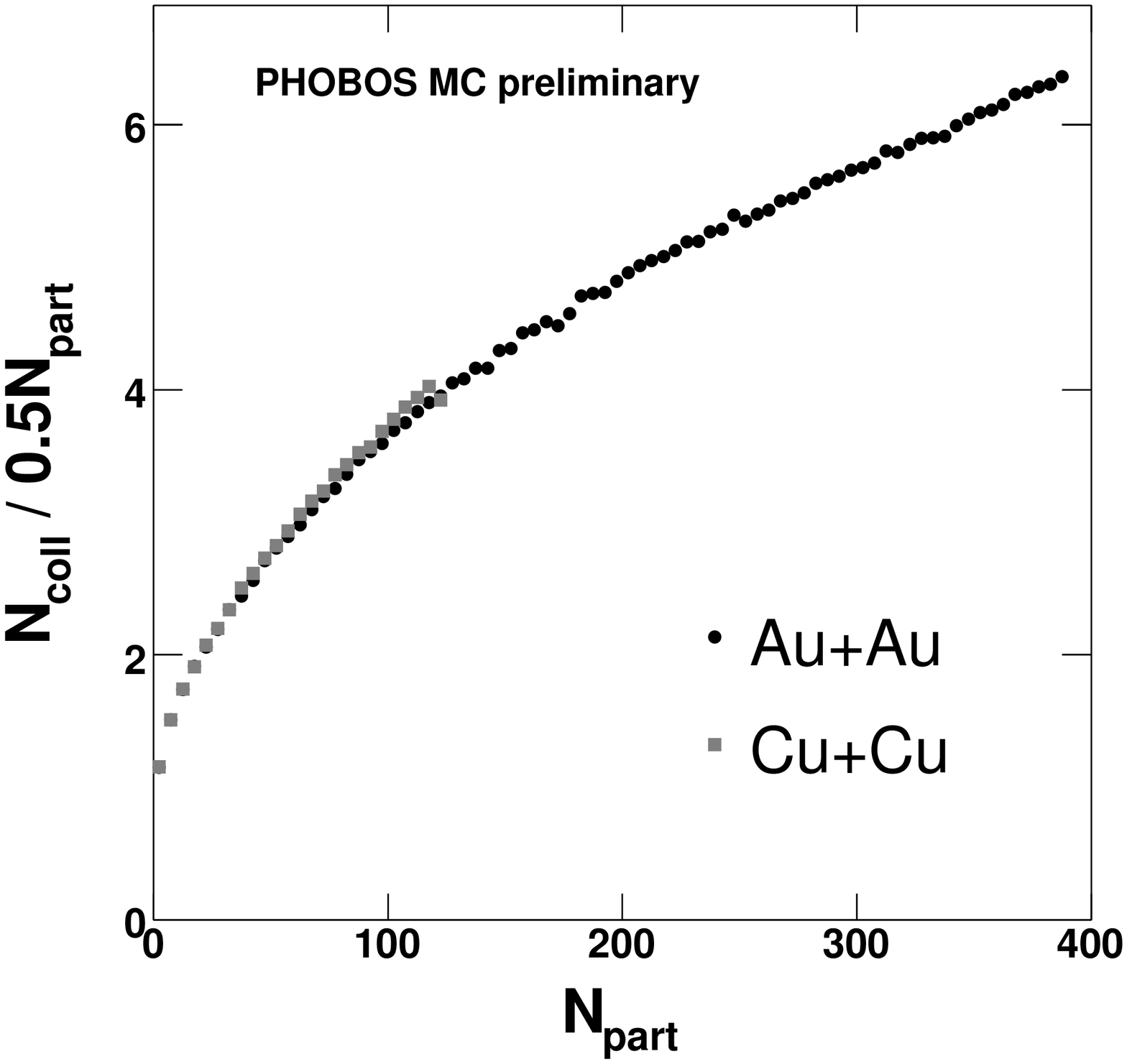}
\caption{Ratio of the average number of collisions per participant, $\bar{\nu}$, for Cu+Cu (light symbols) and 
Au+Au (dark symbols) as a function of $N_{part}$ for $\sqrt{s_{_{NN}}} = 200$~GeV from PHOBOS Glauber MC.}
\label{cucu_auau_nubar}
\end{minipage}
\hspace{\fill}
\begin{minipage}[t]{75mm}
\includegraphics[width=7cm]{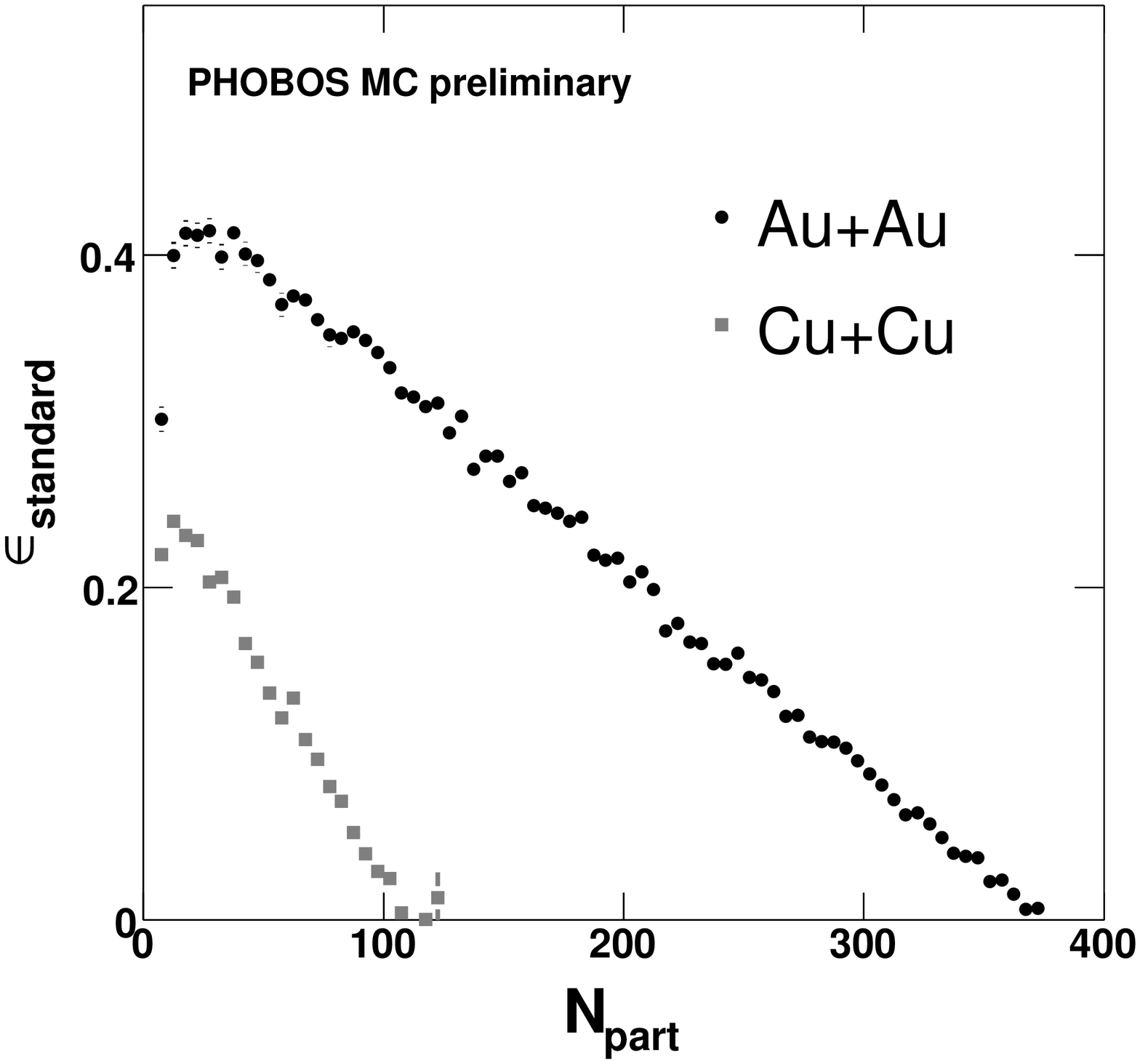}
\caption{Average eccentricity, $\epsilon_{std}$, of the collision zone in Cu+Cu (light symbols) and Au+Au (dark symbols) as 
a function of $N_{part}$ for $\sqrt{s_{_{NN}}} = 200$~GeV from PHOBOS Glauber MC.}
\label{std_ecc_npart}
\end{minipage}
\end{figure}

\begin{table}[h]
\begin{center}
\begin{tabular}{|c|c|c|c|c|}
\hline
                   &{\bf p+p}&{\bf d+Au} & {\bf Cu+Cu} & {\bf Au+Au} \\
\hline
{\bf 410 }GeV       &  20     &           &             &             \\
\hline
{\bf 200 }GeV       &  100    &    150    &       400   &   250       \\
\hline
{\bf 130 }GeV       &         &           &        110  &   4.3       \\
\hline
{\bf 62.4 }GeV      &         &           &             &    22       \\
\hline
{\bf 55.9 }GeV      &         &           &             &   1.8       \\
\hline
{\bf 22.5 }GeV      &         &           &       20    &             \\
\hline
{\bf 19.6 }GeV      &         &           &             &   $\sim1$     \\
\hline
\end{tabular}
\caption{\label{table1} Number of events stored to tape in millions for different systems
from p+p to Au+Au and different collision energies $\sqrt{s_{_{NN}}}$.}
\end{center}
\end{table}

\section{Physics Results}

From the very beginning, the goal of PHOBOS was to obtain a broad survey of particle production in nuclear collisions
for a large variety of system size, collision centralities and collision energies. As Table~\ref{table1} shows, we have 
indeed collected the datasets to fulfill this original program. Using these data, we attempt to extract
organizing principles or scaling features directly from a systematic study of the available observables. As we will show in the 
following, several striking scaling rules have indeed emerged from the data. Some of those appear to connect the results 
obtained in nucleus-nucleus collisions with those in more elementary collisions, like p+p, p+A or $e^+e^-$. Other results show the 
existence of unique collective effects not present in elementary systems. Clearly, the true significance of these features will
be established when they can be shown to naturally emerge from the eventual theoretical understanding of the physics of these
collisions, or when future experiments at much higher energy at the LHC can be shown to be constrained by the same underlying 
features as data at present energies. 

\begin{figure}[t]
\begin{center}
\includegraphics[width=14cm]{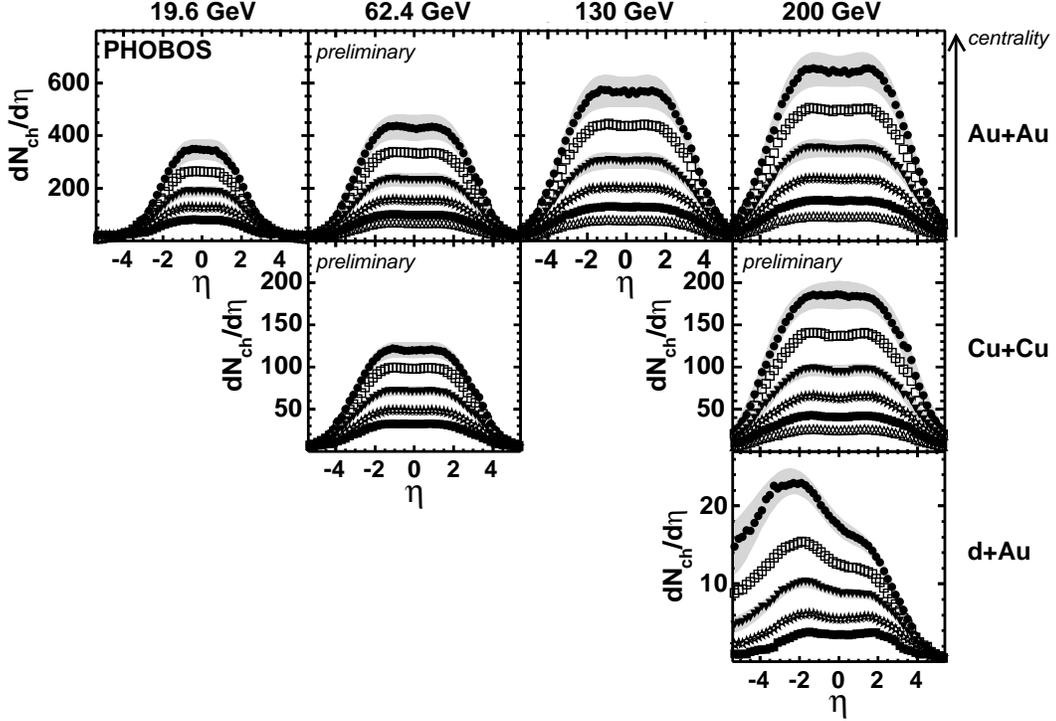}
\end{center}
\vspace{-1.0cm}
\caption{\label{mult_overview}Charged hadron pseudorapidity distributions for different centralities
for Au+Au collisions at $\sqrt{s_{_{NN}}} =$~ 19.6, 62.4, 130 and 200~GeV (top row from left to right), 
Cu+Cu at 62.4 and 200 GeV (middle row, preliminary) and d+Au at 200~GeV (bottom row). }
\end{figure}

In the following subsections, we will use the PHOBOS data to address three distinct questions about scaling of 
particle multiplicities and distributions in nucleus-nucleus collisions. The first question concerns the scaling
of overall particle production in Cu+Cu and Au+Au collisions. In the following section, we will investigate the 
connection between the centrality and energy dependence of particle production and finally we will study the 
connection between of the final-state azimuthal anisotropy and the initial state geometry.

As we will show in the following, conclusive answers emerge from the data for all these questions. 
It is important however to consider what differences between Au+Au and Cu+Cu could be expected based on 
the underlying differences in collision geometry. For this, the result of our Glauber calculations plotted in 
Figs.~\ref{cucu_auau_nubar} and \ref{std_ecc_npart} are important. Shown
in Fig.~\ref{cucu_auau_nubar} is the average number of binary NN collisions $\bar{\nu}$ every nucleon 
undergoes within a nucleus-nucleus collision, as a function of the number of participating nucleons $N_{part}$. 
As this plot shows,
a study of the centrality dependence of particle production allows a large variation of $\bar{\nu}$. However, for a given
value of $N_{part}$, $\bar{\nu}$ in Cu+Cu and Au+Au is the same within the possible experimental resolution and therefore
the Cu+Cu vs.\ Au+Au comparison does not provide additional discriminating power for distinguishing $N_{coll}$ vs.\ 
$N_{part}$ scaling. To better discriminate effects scaling with the number of binary collisions $N_{coll}$ from 
those scaling with $N_{part}$, a comparison of Au+Au results with those from much smaller nuclei like e.g.\ silicon or 
carbon would be necessary. However, Fig.~\ref{std_ecc_npart} shows that there is a very important difference between the 
collision geometry in Cu+Cu and Au+Au, even at the same $N_{part}$. In this figure, the average eccentricity, $\epsilon_{std}$, of 
the collision zone in the transverse plane is plotted against $N_{part}$. Naturally, for the same $N_{part}$ the 
eccentricity in peripheral Au+Au
collisions is much larger than in the corresponding more central Cu+Cu collisions. The comparison of Cu+Cu and Au+Au
will therefore allow a direct investigation of the connection between initial density and the efficiency with which the
collision system translates the initial state geometrical asymmetry, characterized by $\epsilon_{std}$, into a final state 
asymmetry in the momemtum distributions. The results of this comparison, which require a detailed understanding of the definition
of the appropriate eccentricity measure, will be discussed in detail in section~\ref{flow_chapter}.

\begin{figure}[t]
\begin{minipage}[t]{80mm}
\includegraphics[width=7cm]{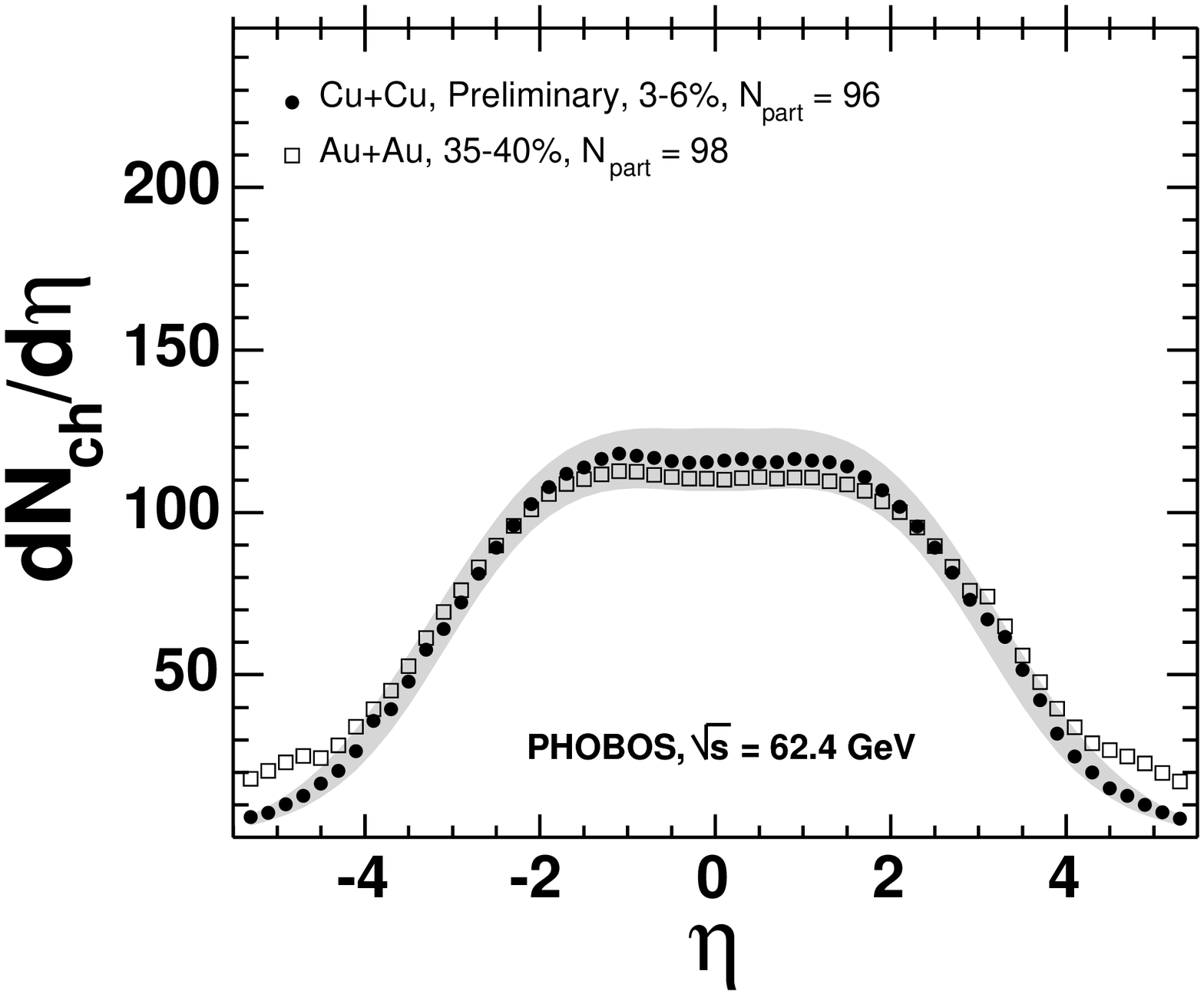}
\caption{Pseudorapidity distribution for charged hadrons in Cu+Cu collisions (black symbols) and Au+Au 
collisions (open symbols) at $\sqrt{s_{_{NN}}} = 62.4$~GeV. The Cu+Cu and Au+Au centralities were selected
to yield similar $N_{part}$. The grey band indicates the systematic uncertainty for Cu+Cu (90\% C.L.). Errors for Au+Au
are not shown.}
\label{cucu_auau_dndeta_62GeV}
\end{minipage}
\hspace{\fill}
\begin{minipage}[t]{75mm}
\includegraphics[width=7cm]{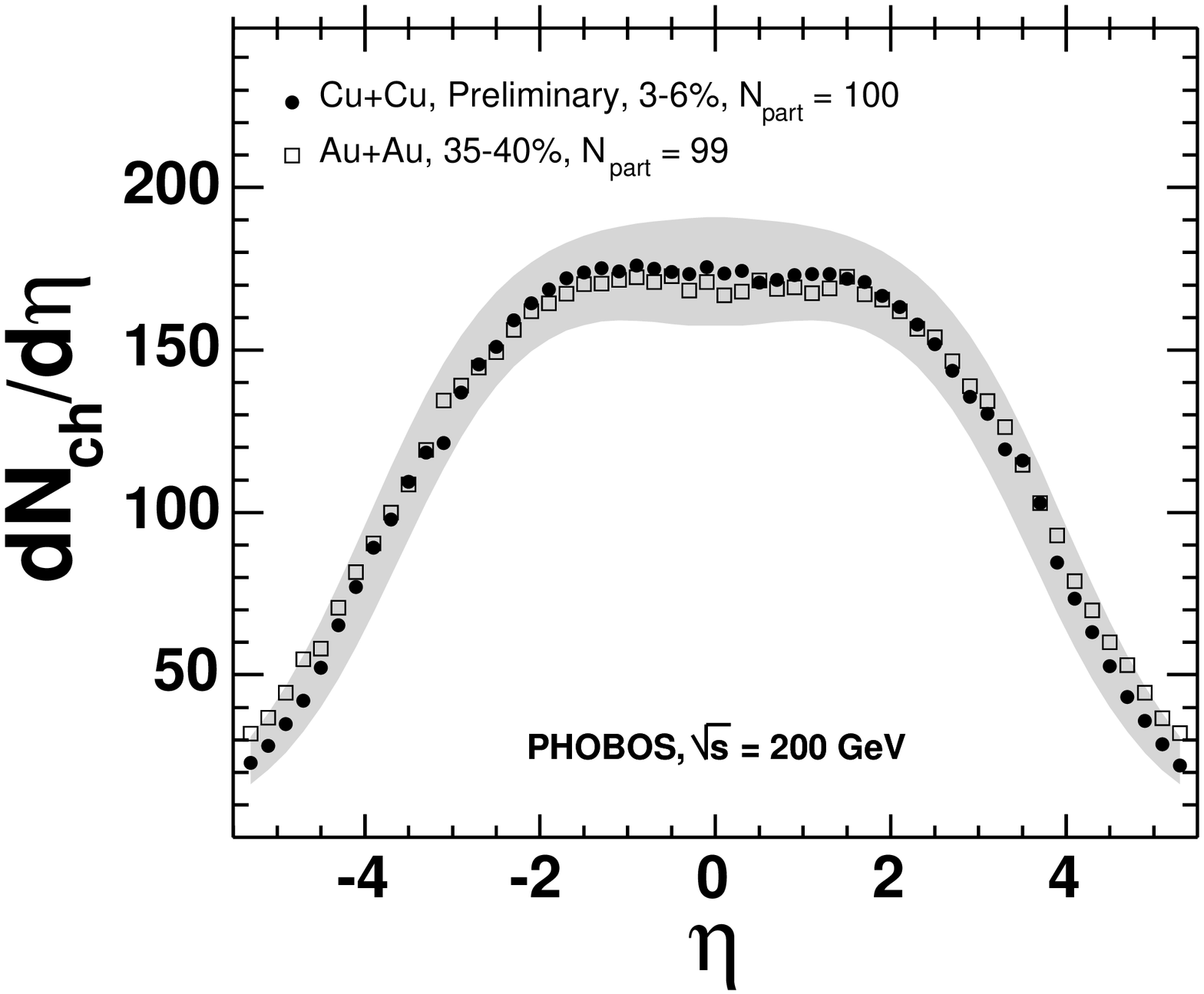}
\caption{Pseudorapidity distribution for charged hadrons in Cu+Cu collisions (black symbols) and Au+Au 
collisions (open symbols) at $\sqrt{s_{_{NN}}} = 200$~GeV. The Cu+Cu and Au+Au centralities were selected
to yield similar $N_{part}$. The grey band indicates the systematic uncertainty for Cu+Cu (90\% C.L.). Errors for Au+Au
are not shown.}
\label{cucu_auau_dndeta_200GeV}
\end{minipage}
\end{figure}

\subsection{System-size dependence of particle production} 

One of the main questions for the Cu+Cu run at RHIC in early 2005 was that of the system-size dependence of 
particle production in nuclear collision at RHIC energies, in terms of overall multiplicity, $dN/d\eta$ distributions 
and $dN/dp_T$ distributions.  Before discussing in detail some of the scaling features observed in the data, 
Fig.~\ref{mult_overview} gives an impression of the breadth and quality of the data collected by PHOBOS. 
The figure shows charged hadron pseudorapidity distributions as a function of centrality
for Au+Au collisions at $\sqrt{s_{_{NN}}} =$~ 19.6, 62.4, 130 and 200~GeV \cite{phobos_limfrag,phobos_mult_62GeV}, 
Cu+Cu at 62.4 and 200 GeV and d+Au at 200~GeV \cite{phobos_dAu_mult}. 

\begin{figure}[t]
\begin{center}
\includegraphics[width=12cm]{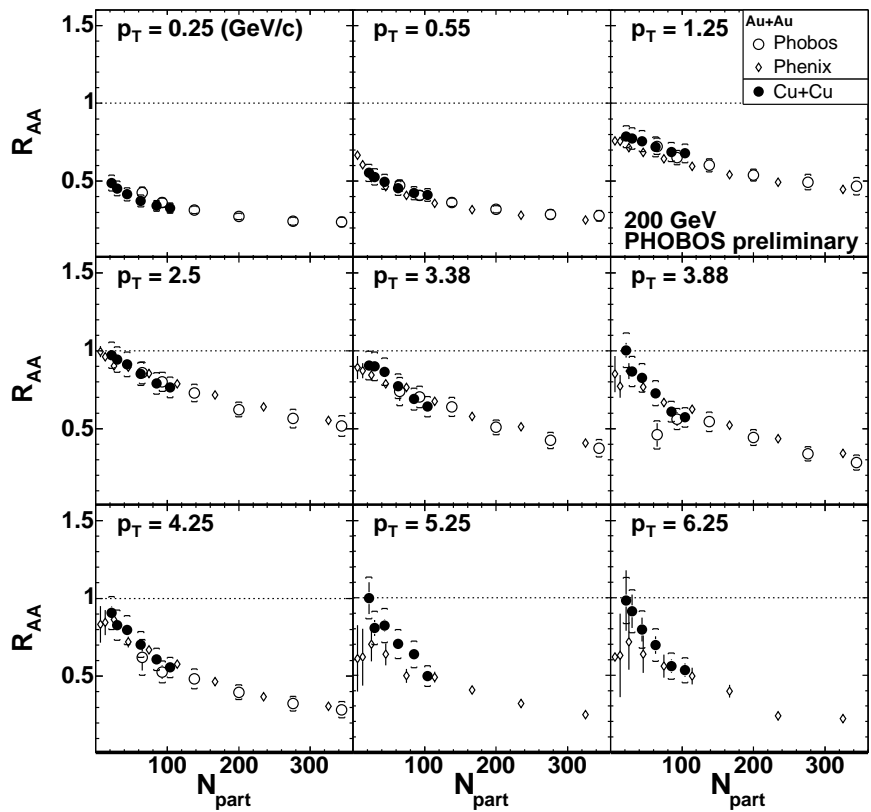}
\vspace{-0.5cm}
\caption{Nuclear modification factor $R_{AA}$ as a function of $N_{part}$ in bins of $p_T$ for 
collisions at$\sqrt{s_{_{NN}}} = 200$~GeV near mid-rapdity. Data for Cu+Cu collisions (filled symbols) are shown in 
comparison with data for Au+Au collisions from PHOBOS (open circles) and PHENIX (open diamonds). 
Systematic errors for Cu+Cu data (90\% C.L.) are shown as brackets. }
\label{raa_cucu_auau_npart}
\end{center}
\end{figure}

For both Au+Au and Cu+Cu collisions, the centrality selection was based on the distribution of energy deposited 
in the region of the multiplicity counter covering $|\eta| < 3.0$ and in the trigger counters covering $3.2 < |\eta| < 4.5$.
The energy distributions for a minimum bias dataset are divided into bins of fractional cross-section and the 
average number of participants for each bin is determined from a Glauber calculation using the HIJING event 
generator \cite{hijing} and a GEANT-based simulation of the PHOBOS detector response. Details of this 
method can be found in \cite{phobos_whitepaper}. 

For the comparison of $dN/d\eta$ distributions in Cu+Cu and Au+Au, we selected centrality bins in Cu+Cu and 
Au+Au such that the average number of participants approximately matched. 
The result of this comparison is shown in Figs.~\ref{cucu_auau_dndeta_62GeV} and \ref{cucu_auau_dndeta_200GeV}
for collisions at 62.4~GeV and 200~GeV, respectively. No further scaling factors are applied. This comparison, and further studies 
for different centrality bins, provide a strikingly simple answer to the question above: If Cu+Cu and Au+Au collisions
at the same collision energy are selected to have the same $\langle N_{part} \rangle$, the resulting 
charged particle $dN/d\eta$ distributions are nearly identical, both in the mid-rapidity particle density 
and the width of the distribution. This is true for both 62.4~GeV and 200~GeV data.

Following the similarity in both the mid-rapidity density and the shape of the pseudorapidity distribution, it is 
natural to compare particle production as a function of transverse momentum in Cu+Cu and Au+Au, for different centrality classes.
This comparison can be seen in Fig.\ref{raa_cucu_auau_npart}, where we plot the nuclear modification factor $R_{AA}$ as 
a function of $N_{part}$ for Cu+Cu and Au+Au collisions at 200~GeV, in bins of $p_T$ from 0.25 to 6.25 GeV/c.  $R_{AA}$ 
is defined as
\begin{equation}
R_{\it AA}(p_T) = \frac{\sigma_{pp}^{inel}}{\langle N_{\it coll} \rangle}
              \frac{d^2 N_{\it AA}/dp_T d\eta} {d^2 \sigma_{pp}/dp_T d\eta}.
\end{equation}
For Au+Au,
data from PHOBOS \cite{phobos_spectra_200GeV} and PHENIX \cite{phenix_spectra_200GeV} are shown. It is important to note 
again that the same $N_{part}$ value in Cu+Cu and Au+Au corresponds to virtually the same value of $N_{coll}$ for both systems. 
Therefore, the fact that $R_{AA}$ coincides for Cu+Cu and Au+Au at the same $N_{part}$, as shown in Fig.~\ref{raa_cucu_auau_npart}
implies that for a given system size, measured by either $N_{part}$ or $N_{coll}$, the absolute yield per participant 
is the same in both systems. This statement holds over the entire range in centrality and $p_T$ covered in this analysis
and is also confirmed with similar precision in collisions at 62.4 GeV.
In summary, overall particle production per participant, as well as particle distribution in pseudorapidity and transverse 
momentum, appear identical in Cu+Cu and Au+Au at a given collision energy, 
if collisions with the same average number of participants are selected.

\subsection{Factorization of energy and centrality dependence}

\begin{figure}[t]
\begin{minipage}[t]{80mm}
\includegraphics[width=8cm]{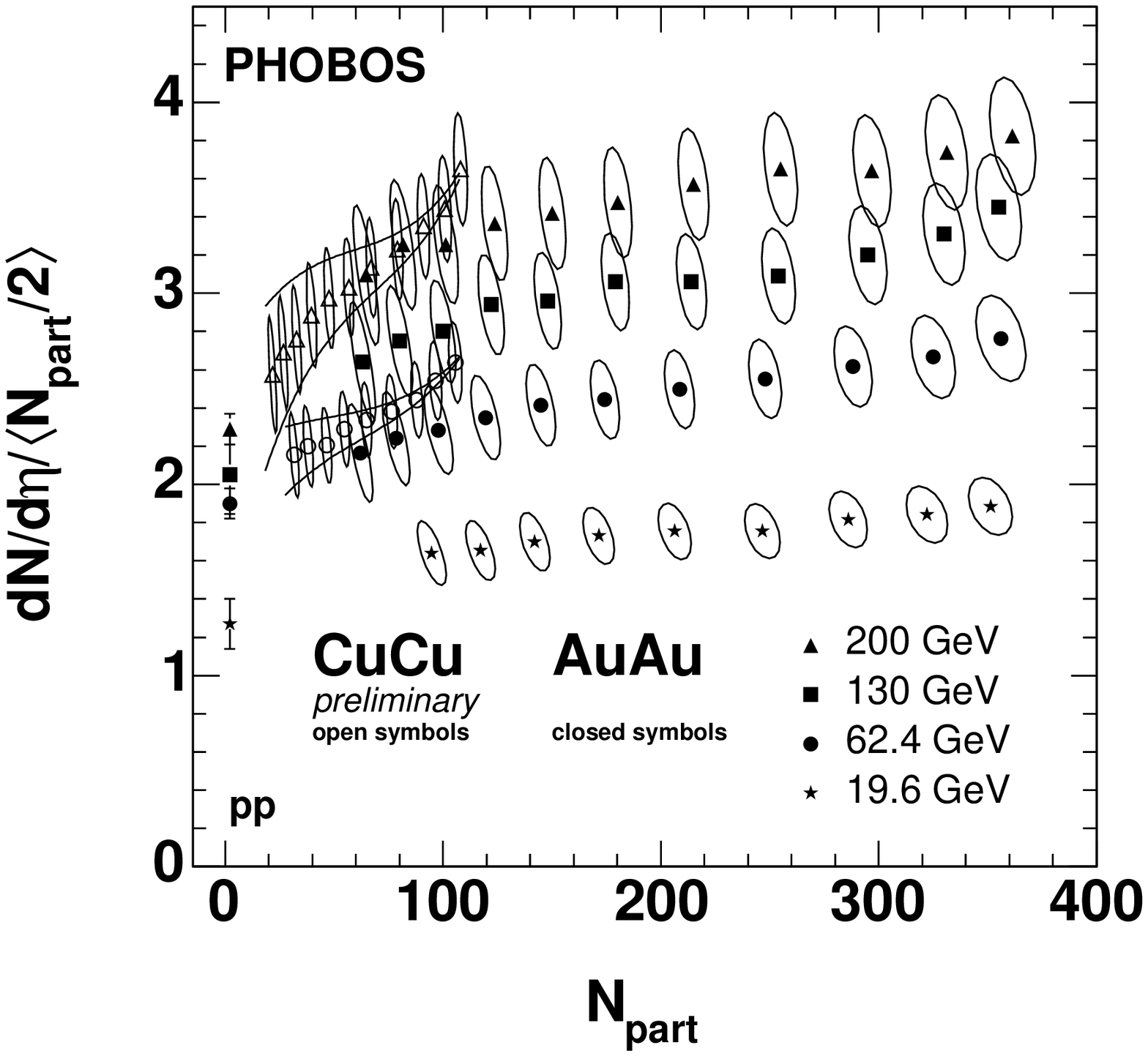}
\vspace{-0.5cm}
\caption{Mid-rapidity density $dN/d\eta / \langle N_{part}/2 \rangle$ as a function of $N_{part}$. Au+Au data for
$\sqrt{s_{_{NN}}} =$~ 19.6, 62.4, 130 and 200~GeV are shown as filled symbols, Cu+Cu data for $\sqrt{s_{_{NN}}} =$~ 62.4,
and 200~GeV are shown as open symbols. Ellipses indicate the uncertainty in $N_{part}$ and $dN/d\eta$ determination. Lines indicate
the additional uncertainty due to the uncertainty in the trigger cross-section for Cu+Cu.}
\label{mid-rap_dndeta_npart}
\end{minipage}
\hspace{\fill}
\begin{minipage}[t]{75mm}
\includegraphics[width=8cm]{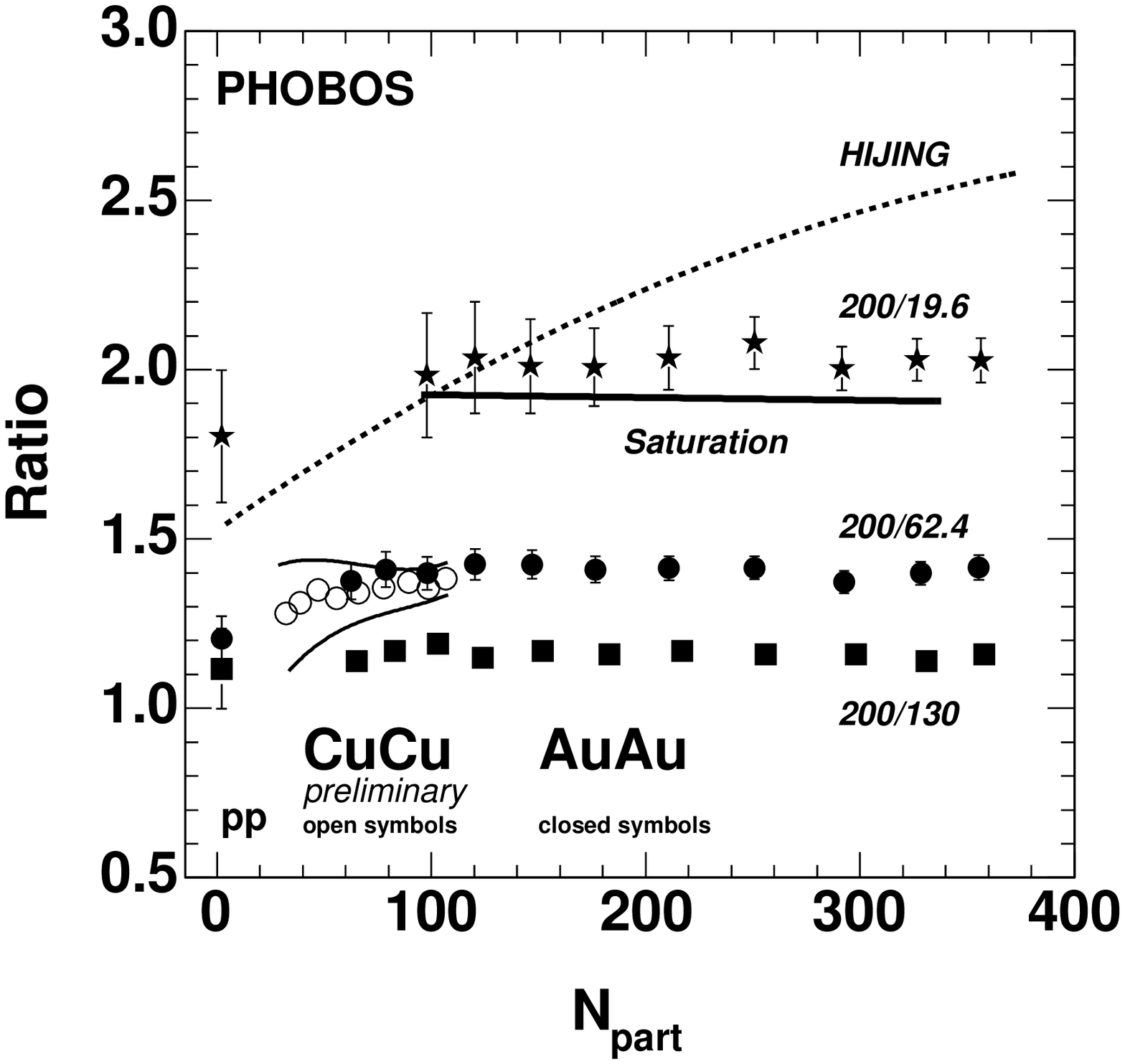}
\vspace{-0.5cm}
\caption{Ratio of mid-rapidity densities $dN/d\eta / \langle N_{part}/2 \rangle$ as a function of $N_{part}$. Au+Au data for
energies 200/19.6, 200/62.4, 200/130 (filled symbols) and Cu+Cu data for 200/62.4 (open symbols) are shown, 
in comparison to HIJING calculations and predictions from Kharzeev et al.\ for the 200/19.6 ratio \cite{kharzeev}. 
Lines indicate the additional uncertainty due to the 
uncertainty in the trigger cross-section for Cu+Cu. Ratios for p+p collisions are shown at $N_{part}=2$.}
\label{mid-rap_ratio_npart}
\end{minipage}
\end{figure}

While the comparison of particle production in Cu+Cu and Au+Au at the same collision energy yields the simplest possible
connection between the two systems, a simultaneous study of centrality and energy dependence of the same observables for
either of the systems yields a subtle and surprising scaling relationship, that again holds for mid-rapidity particle production,
the shape of the $dN/d\eta$ distributions and the transverse momentum distributions. As was first shown for the mid-rapidity 
density \cite{phobos_tracklets}, the observed increase in particle production per participant with increasing $N_{part}$ 
(see Fig.\ref{mid-rap_dndeta_npart}) is independent of collision energy over the full energy range of RHIC from 20 to 200~GeV.

This is illustrated in Fig.\ref{mid-rap_ratio_npart}, where we show the ratio of mid-rapidity densities as a function of 
$N_{part}$ relative to 200~GeV data. All ratios, from 200/130~GeV to 200/19.6~GeV, are flat within the experimental uncertainty. 
This factorization of energy and centrality dependence is strongly violated in models that describe overall particle production as a superposition of independent ``soft'' contributions scaling like $N_{part}$ and ``hard'' contributions scaling like $N_{coll}$,
where the increase in the hard contributions is given by the increasing mini-jet production with energy. As a quantitative 
example, Fig.~\ref{mid-rap_ratio_npart} shows the prediction from HIJING. It is interesting to note that the
observed factorization is described in approaches based on the ideas of parton saturation \cite{kharzeev,wiedemann_mult}.

\subsubsection{Energy and centrality factorization of $p_T$ distributions}

\begin{figure}[t]
\begin{center}
\includegraphics[width=15.5cm]{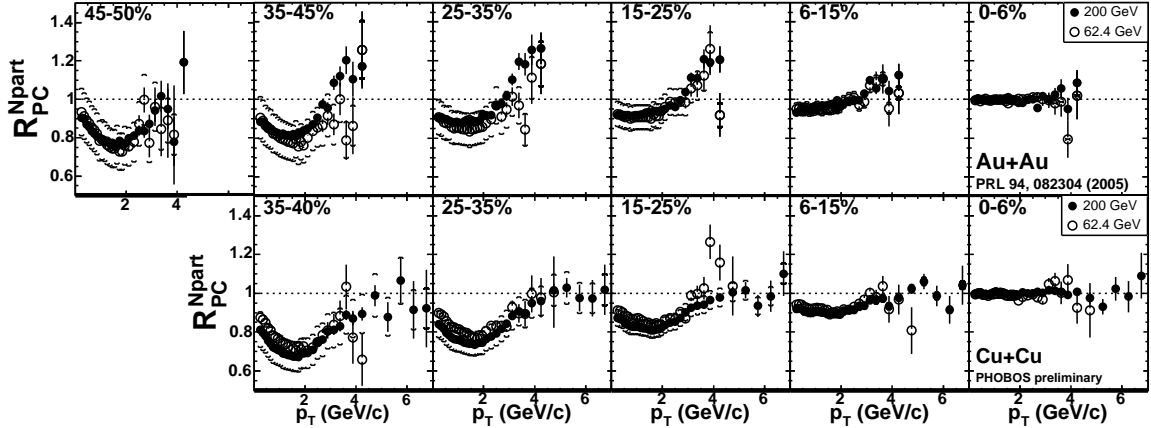}
\end{center}
\vspace{-1.0cm}
\caption{\label{ratio_spectra_pT} Ratios of yields per participant vs $p_T$ (near mid-rapidity), 
relative to central collisions for bins of 
fractional  cross-section. Central events are shown in the right-most panel. Solid symbols show 200 GeV data, open symbols 
show 62.4 GeV data. Top row shows Au+Au data, lower row shows Cu+Cu data. Systematic uncertainties (90\% C.L.) are shown by brackets.}
\end{figure}

In an earlier publication \cite{phobos_62GeV_spectra}, we showed that the factorization described above not only holds for the 
$p_T$-integrated yields at mid-rapidity, but also differentially over a large range of transverse momenta. This is illustrated in 
the top row of Fig.~\ref{ratio_spectra_pT} for Au+Au collisions at 62.4 and 200 GeV. There, the quantity $R_{\it PC}^{N_{part}}(p_T)$, 
defined as
\begin{equation}
R_{\it PC}^{N_{part}}(p_T) = \frac{\langle N_{\it part}^{0-6\%} \rangle}{\langle N_{\it part} \rangle}
              \frac{d^2 N_{\it AA}/dp_T d\eta} {d^2 N_{\it AA}^{0-6\%}/dp_T d\eta},
\end{equation}
is shown as a function of $p_T$ for the six centrality bins. Remarkably, the centrality evolution of the spectral shape
from central to peripheral collisions is found to agree within the experimental uncertainty between 62.4 and 200~GeV,
even though $R_{AA}$ itself shows a large variation by more than a factor of two at high $p_T$ between the two energies.
The bottom row of Fig.~\ref{ratio_spectra_pT} shows that the same energy-independence of the shape evolution is seen for the
much smaller Cu+Cu system, when again plotting $R_{PC}(p_T)$ in bins of fractional cross-section. This apparent dominance of collision
geometry, relative to expected dynamical effects, is an important feature of the data that remains to be understood.

\subsubsection{Extended longitudinal scaling}

\begin{figure}[t]
\begin{minipage}[t]{80mm}
\includegraphics[width=5cm]{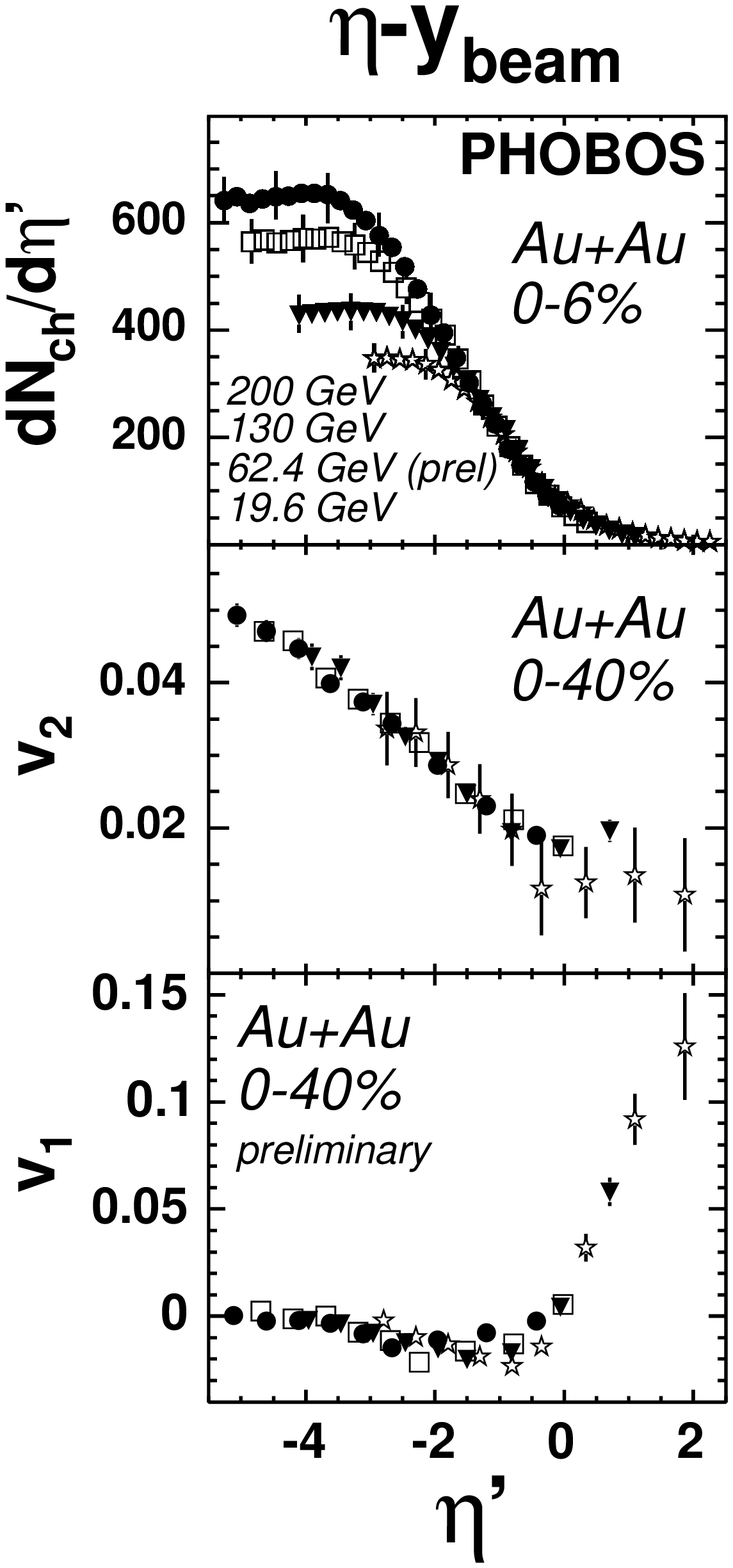}
\caption{\label{limfrag_auau} Dependence of $dN/d\eta'$ (top), elliptic flow $v_2$ (middle) and directed flow $v_1$ 
(bottom) on $\eta' = |\eta| - y_{beam}$ for Au+Au collisions at $\sqrt{s_{_{NN}}} =$~ 19.6, 62.4, 130 and 200~GeV.}
\end{minipage}
\hspace{\fill}
\begin{minipage}[t]{75mm}
\includegraphics[width=5cm]{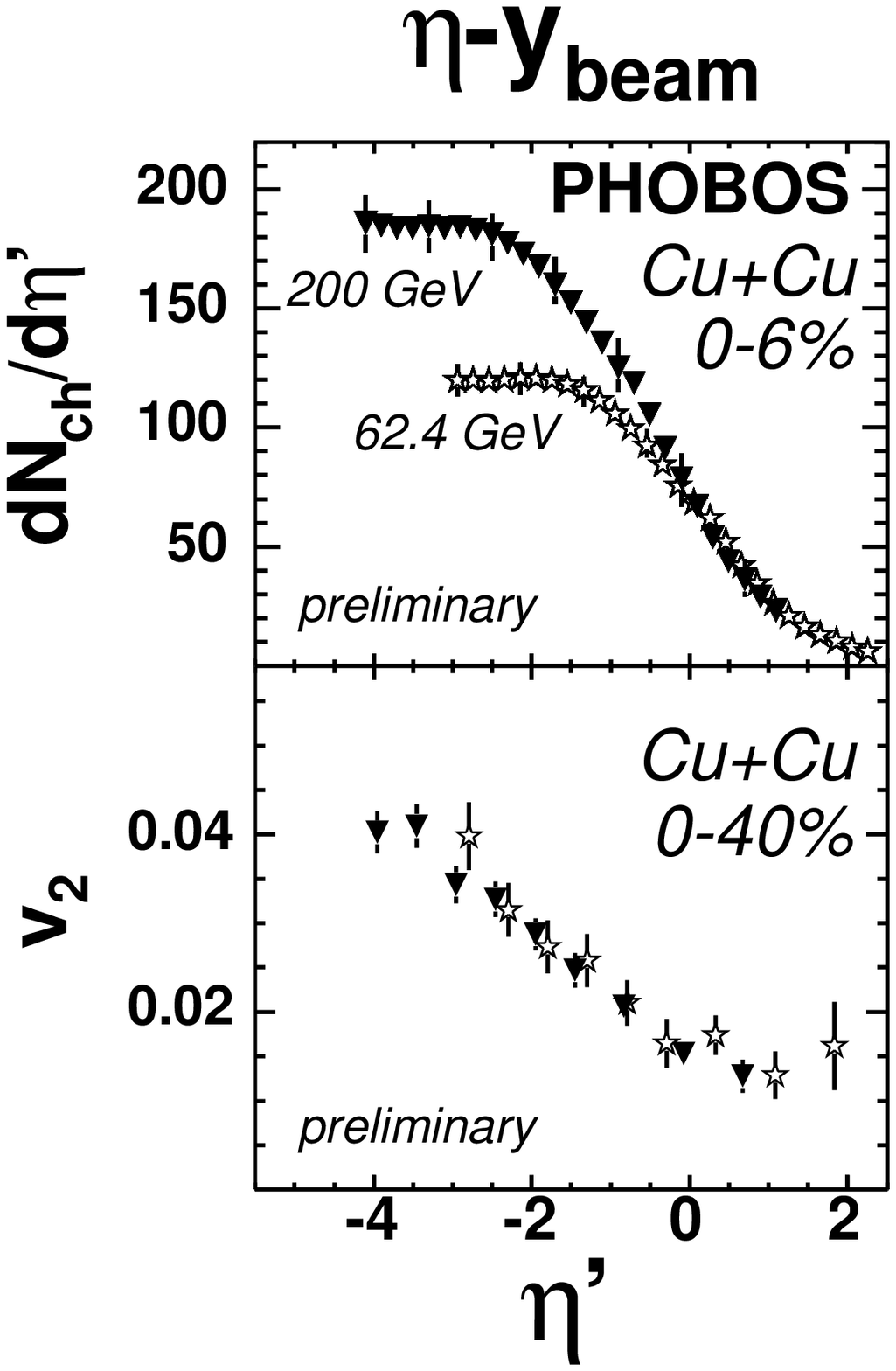}
\caption{\label{limfrag_cucu} Dependence of $dN/d\eta'$ (top) and  elliptic flow $v_2$  
(bottom) on $\eta' = |\eta| - y_{beam}$ for Cu+Cu collisions at $\sqrt{s_{_{NN}}} =$~ 62.4 and 200~GeV (preliminary).}
\end{minipage}
\end{figure}

As shown in earlier publications, the energy and centrality dependence of particle production factorize not only for 
mid-rapidity multiplicity densities. One of the most striking scaling relationships seen in nuclear collisions at RHIC
is the energy-independence of particle yields at moderate to high rapidities, when viewed in the restframe of one of the colliding 
nuclei. This phenomenon has also been observed in p+p and p+A collisions over a large range of energies and is commonly referred to
as ``limiting fragmentation''. PHOBOS has extended this observation not only to $dN/d\eta$ distributions in nuclear collisions 
at different collision centralities \cite{phobos_limfrag}, but also to the pseudorapidity dependence of azimuthal anisotropies
\cite{phobos_v2_limfrag,tonjes_qm04}. 
In Fig.~\ref{limfrag_auau}, a compilation of results for $dN/d\eta$,  the elliptic flow coefficient $v_2$ 
and the directed flow coefficient $v_1$ is shown for Au+Au collisions at four different collision energies from 20 to 200~GeV. 

As can be seen in this figure, limiting fragmentation holds not only for the multiplicity distribution of charged hadrons,
but also for observables like $v_2$. This is remarkable as $v_2$ is believed to be generated dynamically in the evolution
of the collision, from presumably very different initial conditions at the different collision energies. Furthermore,
Fig.~\ref{limfrag_auau} shows that the scaling behaviour is not confined to a small ``fragmentation region'' near beam-rapidity, 
but extends over a large part of longitudinal phase space. For this reason, we refer to this set of observations as ``extended
longitudinal scaling''. Finally, Fig.~\ref{limfrag_cucu} shows that the same 
scaling behaviour also holds for $dN/d\eta$ and $v_2$ in Cu+Cu. Measurements of $v_1$ in Cu+Cu are forthcoming.

\begin{figure}[t]
\begin{minipage}[t]{80mm}
\includegraphics[width=7cm]{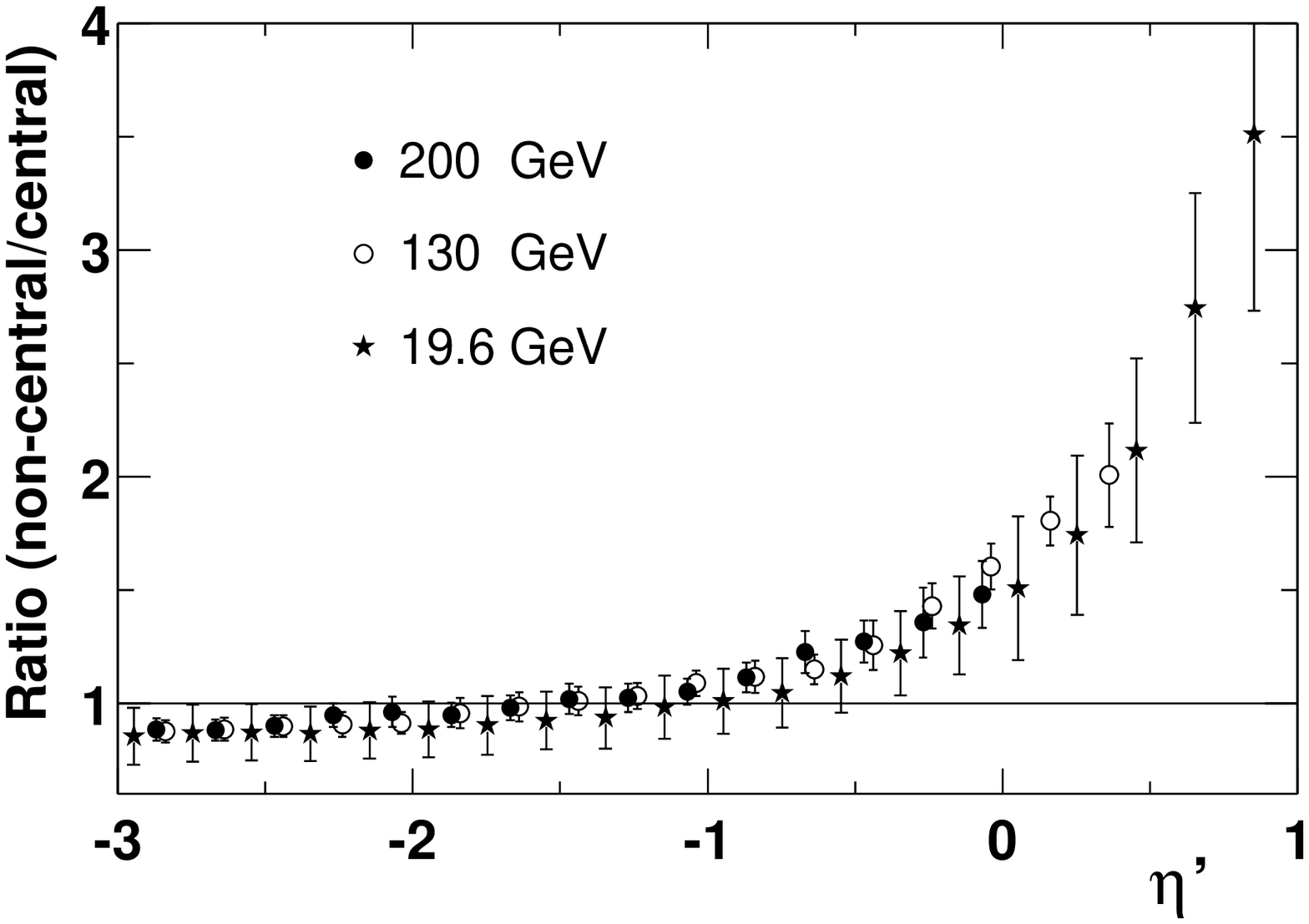}
\caption{Ratio $R_{PC}(\eta)$ of $dN/d\eta'$ distributions for peripheral to central events for Au+Au collisions at
$\sqrt{s_{_{NN}}} =$~ 19.6, 130 and 200~GeV as a function of $\eta' = |\eta| - y_{beam}$. 
Only systematic uncertainties (90\% C.L.) are shown.}
\label{limfrag_rpc}
\end{minipage}
\hspace{\fill}
\begin{minipage}[t]{75mm}
\includegraphics[width=7cm]{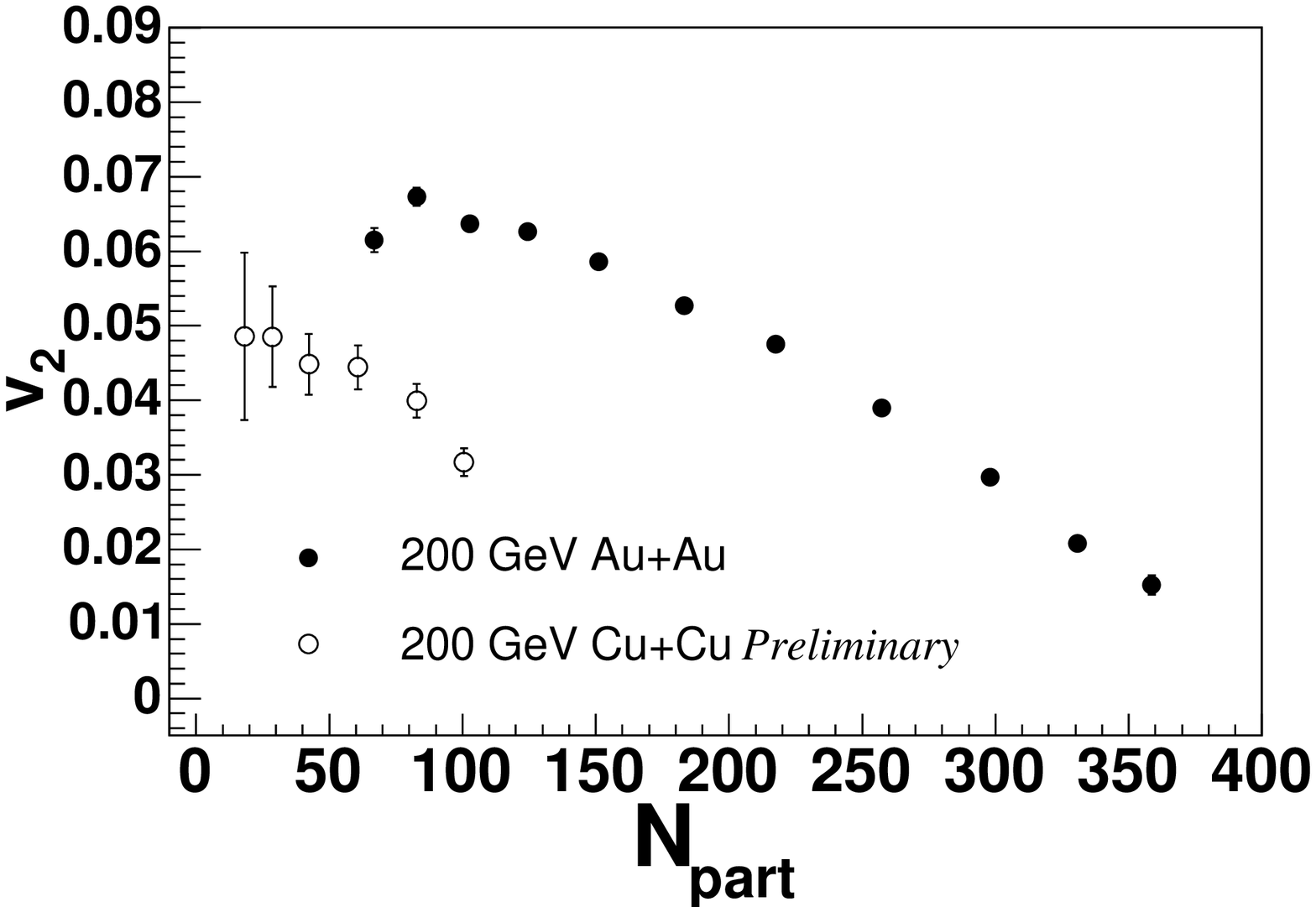}
\caption{Average elliptic flow coefficient $v_2$ measured near mid-rapidity,
 as a function of $N_{part}$ for Cu+Cu collisions (open symbols) and 
Au+Au collisions (filled symbols) at $\sqrt{s_{_{NN}}} =$~ 200~GeV. Only statistical errors are shown.}
\label{v2_cucu_auau_npart}
\end{minipage}
\end{figure}

In order to examine the connection between energy- and centrality-scaling for longitudinal distributions, it is useful to 
again use the ratio of yields for peripheral relative to central events, $R_{PC}^{N_{part}}(\eta)$:
\begin{equation}
R_{\it PC}^{N_{part}}(\eta) = \frac{\langle N_{\it part}^{0-6\%} \rangle}{\langle N_{\it part} \rangle}
              \frac{d N_{\it AA}/d\eta} {d N_{\it AA}^{0-6\%}/d\eta}
\end{equation}
In Fig.~\ref{limfrag_rpc} we show that also the longitudinal distributions, when viewed in the restframe of one 
of the colliding nuclei, exhibit the factorization seen for mid-rapidity yields. 

In summary, the combined study of energy- and system size dependence of hadronic observables in Au+Au collisions has 
revealed a remarkable factorization, leading to an energy-independence of the centrality evolution for 
the mid-rapidity particle density, the charged hadron $p_T$ distributions near mid-rapidity and the extended
longitudinal scaling of multiplicity distributions. This factorization is particularly surprising in the context
of a hard/soft picture of particle production, where the increasing mini-jet cross-sections with energy and 
the increasing ratio of collisions per participant, $\bar{\nu}$, couple energy and centrality dependence.
It appears that the observed dominance of collision geometry, which is naturally energy-independent for a 
given centrality selection, is better captured in a description of the collision process based on ideas 
of parton saturation.

\subsection{\label{flow_chapter} System size dependence of elliptic flow}

The final question to be answered by the comparison of Au+Au and Cu+Cu data as a function of energy
and centrality concerns elliptic flow, and the connection between the initial state conditions and the 
observed final state anisotropy. It has been argued in the past \cite{star_flow_2000}, that the observed 
flow coefficient $v_2$ for Au+Au closely follows the initial state eccentricity as a function of centrality. This 
can be seen by comparing $N_{part}$ dependence for the Au+Au initial state eccentricity from a 
Glauber calculation shown in Fig.~\ref{std_ecc_npart} and the Au+Au $v_2$ coefficient at mid-rapidity shown in 
Fig.~\ref{v2_cucu_auau_npart}. However, doubt on this crucial connection is cast by comparing the Cu+Cu calculation
and data in the same figures. Whereas the average eccentricity for Cu+Cu tends to zero for the most central collisions, 
the corresponding $v_2$ values for Cu+Cu only drop to a value of $v_2 \approx 0.03$ even for the most central 
collisions. This would lead to the paradoxical conclusion that for the same $N_{part}$ and therefore the same
initial area density of produced particles, the Cu+Cu system is much more effective in translating an initial
eccentricity into a final state anisotropy than the Au+Au system. Alternatively, large non-flow effects mimicking 
a dynamically generated anisotropy could be postulated for the Cu+Cu system. 

\begin{figure}[t]
\begin{minipage}[t]{80mm}
\includegraphics[width=7cm]{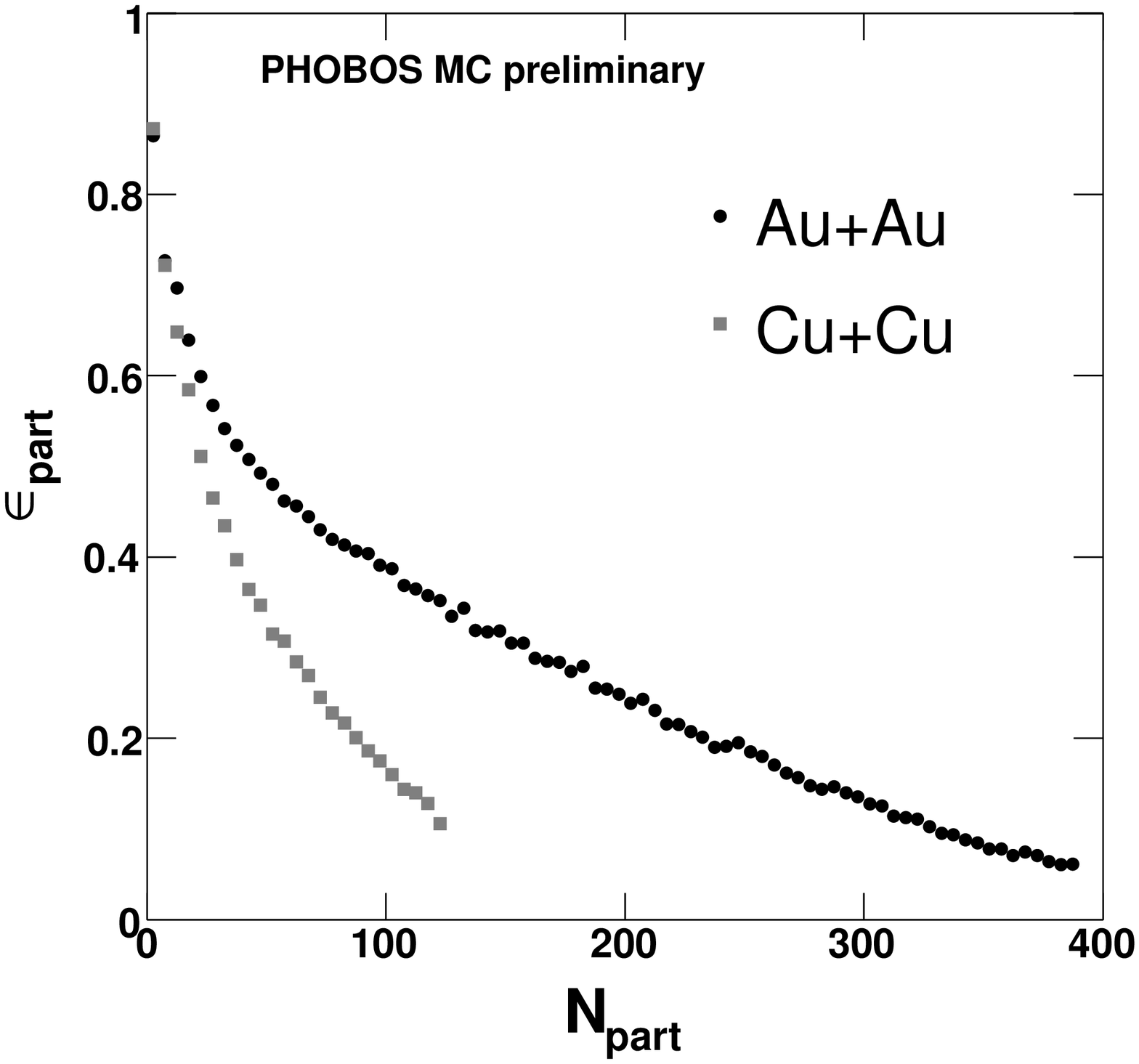}
\caption{Participant eccentricity $\epsilon_{part}$ of the collision zone in Cu+Cu (light symbols) and Au+Au (dark symbols) as
a function of $N_{part}$ for $\sqrt{s_{_{NN}}} = 200$~GeV from PHOBOS Glauber MC.}
\label{ecc_part_cucu_auau}
\end{minipage}
\hspace{\fill}
\begin{minipage}[t]{75mm}
\includegraphics[width=7cm]{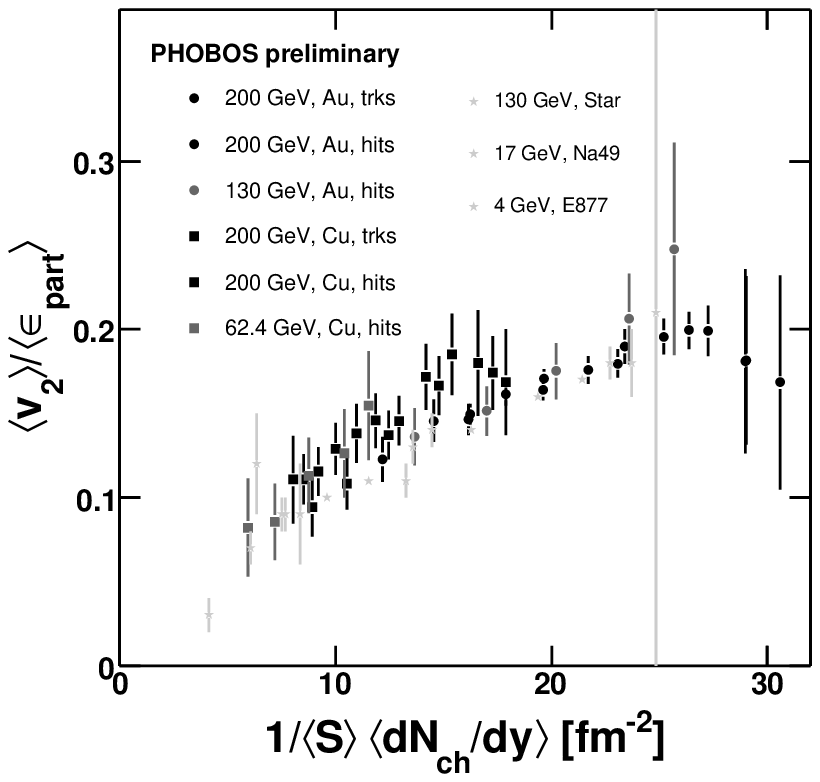}
\caption{Ratio of $v_2$ coefficient to average participant eccentricity as a function of mid-rapidity area density $1/\langle S \rangle 
\langle dN/dy \rangle$ for Au+Au collisions at $\sqrt{s_{_{NN}}} =$~ 130 and 200~GeV and Cu+Cu collisions 
at $\sqrt{s_{_{NN}}} =$~ 62.4  and 200~GeV, in comparison to lower energy data from NA49 and E877. }
\label{ldl_scaling}
\end{minipage}
\end{figure}

However, a possible explanation unifying the observations in Au+Au and Cu+Cu can be found by examining the 
underlying definition of the initial state eccentricity. This eccentricity, called $\epsilon_{std}$ in 
the following, is commonly defined
as the average eccentricity of the distribution of participating nucleons, relative to the known reaction plane, 
obtained for a certain centrality class in a Glauber calculation. This definition suffers from two potential problems:
It averages out the fluctuations from event-to-event in the actual participant distributions. Finite number
fluctuations will lead to an eccentric nucleon distribution even for collisions with impact parameter $b=0$. With the 
standard definition, these fluctuations will be averaged to zero for central events. Furthermore, the minor axis 
for the actual event-by-event participant distribution will in general not coincide with the impact parameter 
vector. The eccentricity calculated relative to the reaction plane will therefore underestimate the true 
eccentricity of the nucleon distribution. To study these deficiencies, we have defined an alternative measure of 
the eccentricty in each centrality bin, where we calculate the eccentricity for each Glauber event relative to the 
principal axes of the actual participant distribution (see \cite{manlyqm05}). 
By construction, this \em participant eccentricity \em, 
$\epsilon_{part}$, will always be positive and will therefore average to a finite value even for the most central 
events. In addition, the smaller number of colliding nucleons, makes the difference 
between $\epsilon_{std}$ and $\epsilon_{part}$ particularly important for the Cu+Cu system relative to Au+Au.
The result of a Glauber calculation for $\epsilon_{part}$ for Cu+Cu and Au+Au as a function of $N_{part}$ can be seen in 
Fig.~\ref{ecc_part_cucu_auau}. As expected, $\epsilon_{part}$  remains finite even for the most central Cu+Cu collisions.

Using $\epsilon_{part}$, we can now attempt to identify a common scaling behaviour of Cu+Cu and Au+Au collisions 
over a large range of collision energies and centralities. This is shown in Fig.~\ref{ldl_scaling}, plotting the ratio
of $\langle v_2 \rangle / \langle \epsilon_{part} \rangle$ versus the mid-rapidity area density of produced particles 
\cite{ldl_voloshin,ldl_heiselberg}. The data appear to exhibit a common scaling behavior over a large 
range in collision energy, suggesting that the efficiency for translating the initial state eccentricity 
estimated using $\epsilon_{part}$ into a final state anisotropy $v_2$ appears to only depend on the initial area
density achieved in the collision. Clearly, it is a fascinating question for future experiments whether this curve
saturates at higher densities or continues to rise.

\section{Future PHOBOS physics program}

The future physics program of PHOBOS will rely on the analysis of the existing datasets. The program is characterized by two
major directions. We will use the unique acceptance of our mid-rapidity spectrometer to study particle production
at the lowest transverse momenta accessible at RHIC, as a function of collision energy, centrality and colliding species. 
This includes studies of $\phi$-meson production at very low $p_T$, which requires an analysis of the full high statistics 
datasets for Cu+Cu and Au+Au. The second class of studies uses the high statistics datasets in combination with the large acceptance
of the multiplicity array to study fluctuations and correlations in particle production. Preliminary results from these studies
are presented in separate contributions to these proceedings \cite{adamqm05,peterqm05,georgeqm05} and will be summarized here
briefly.

\subsection{Particle production at low $p_T$}

The unique setup of the PHOBOS silicon spectrometer and its proximity to the interaction vertex 
allows us to detect charged particles that range out and stop in the front layers of the silicon detector. For the present 
analysis, we have developed an algorithm detecting particles that stop in the fifth silicon layer. Based on the total deposited 
energy and the specific energy loss in the silicon detectors, we can reconstruct the momentum and mass of charged particles. 
This allows us to measure pions, kaons and protons (summed with their anti-particles) for transverse momenta as low as
 30, 90 and 140 MeV/c, respectively. 
For details of the reconstruction algorithm, see \cite{stopping_auau_200}. 

At this conference, new results for the 
centrality dependence of particle production at low $p_T$ in Au+Au collisions at 62.4 and 200~GeV were presented, as well as results 
for minimum bias d+Au collisions at 200~GeV. A detailed discussion of the results can be 
found in \cite{adamqm05}. As an example, Fig.~\ref{dau_blastwave} shows the yield for pions, kaons and protons at low $p_T$ in 
combination with results at higher $p_T$ for minimum bias d+Au collisions at 200 GeV. Both results are fully corrected, including a
correction for feed-down from weak decays. Also shown are the results of a blastwave fit to the higher $p_T$ data, which is found to
underestimate the low $p_T$ points for d+Au, whereas for similar fits in Au+Au collisions consistency between
the low and high $p_T$ datasets is found. It is interesting to note however that the effective expansion velocity extracted 
from the fit to the d+Au high $p_T$ data is 0.35~$c$, indicating that the physical interpretation of the extracted value has 
to be treated carefully.

\subsection{Fluctuations and Correlations}

\begin{figure}[t]
\begin{minipage}[t]{80mm}
\includegraphics[width=6cm]{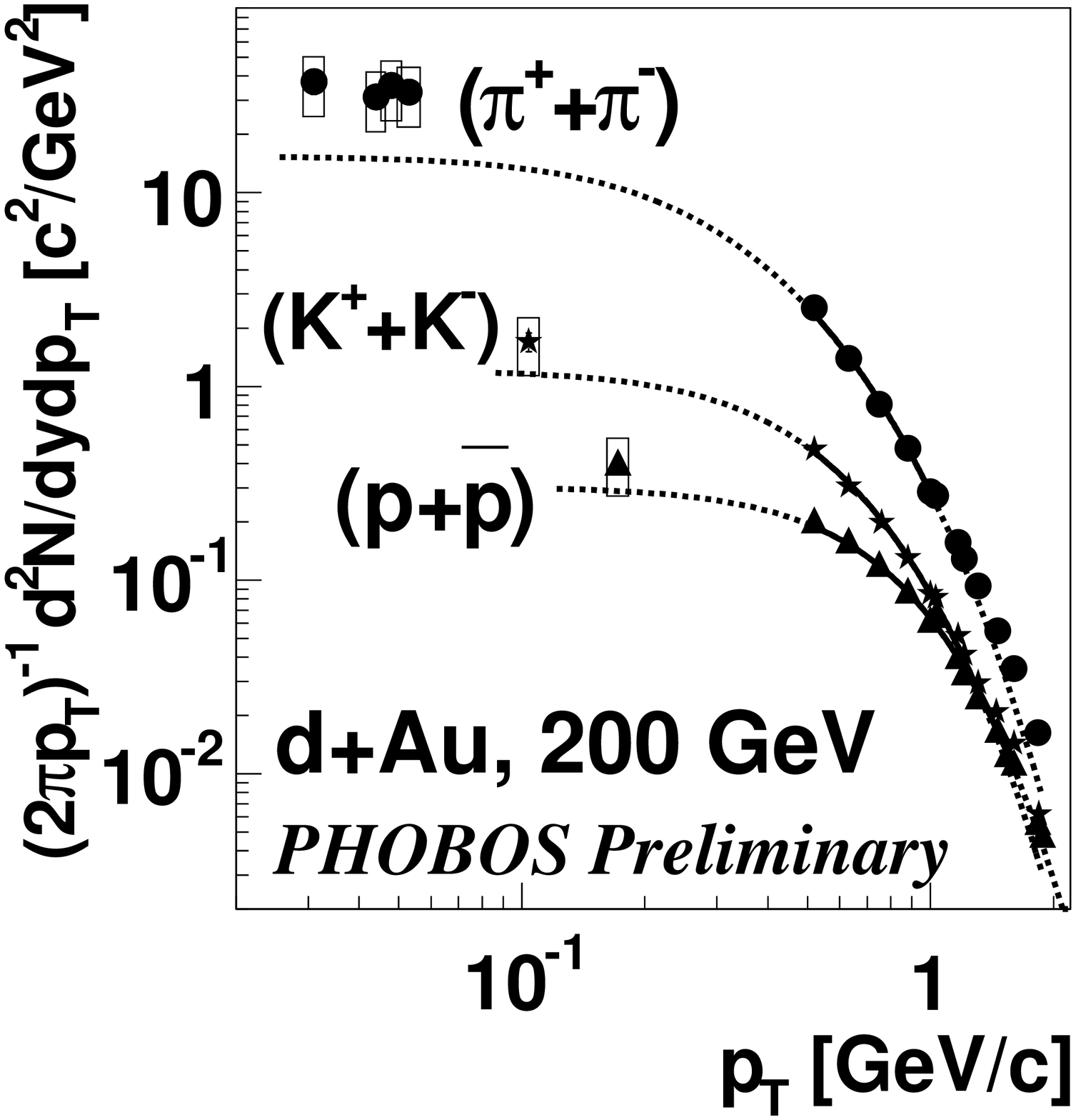}
\caption{Transverse momentum spectra for pions, kaons and protons in 200~GeV min-bias d+Au collisions. Lines indicate
the results of a blast-wave fit for data above $p_T =0.25$~GeV/c.}
\label{dau_blastwave}
\end{minipage}
\hspace{\fill}
\begin{minipage}[t]{75mm}
\includegraphics[width=8cm]{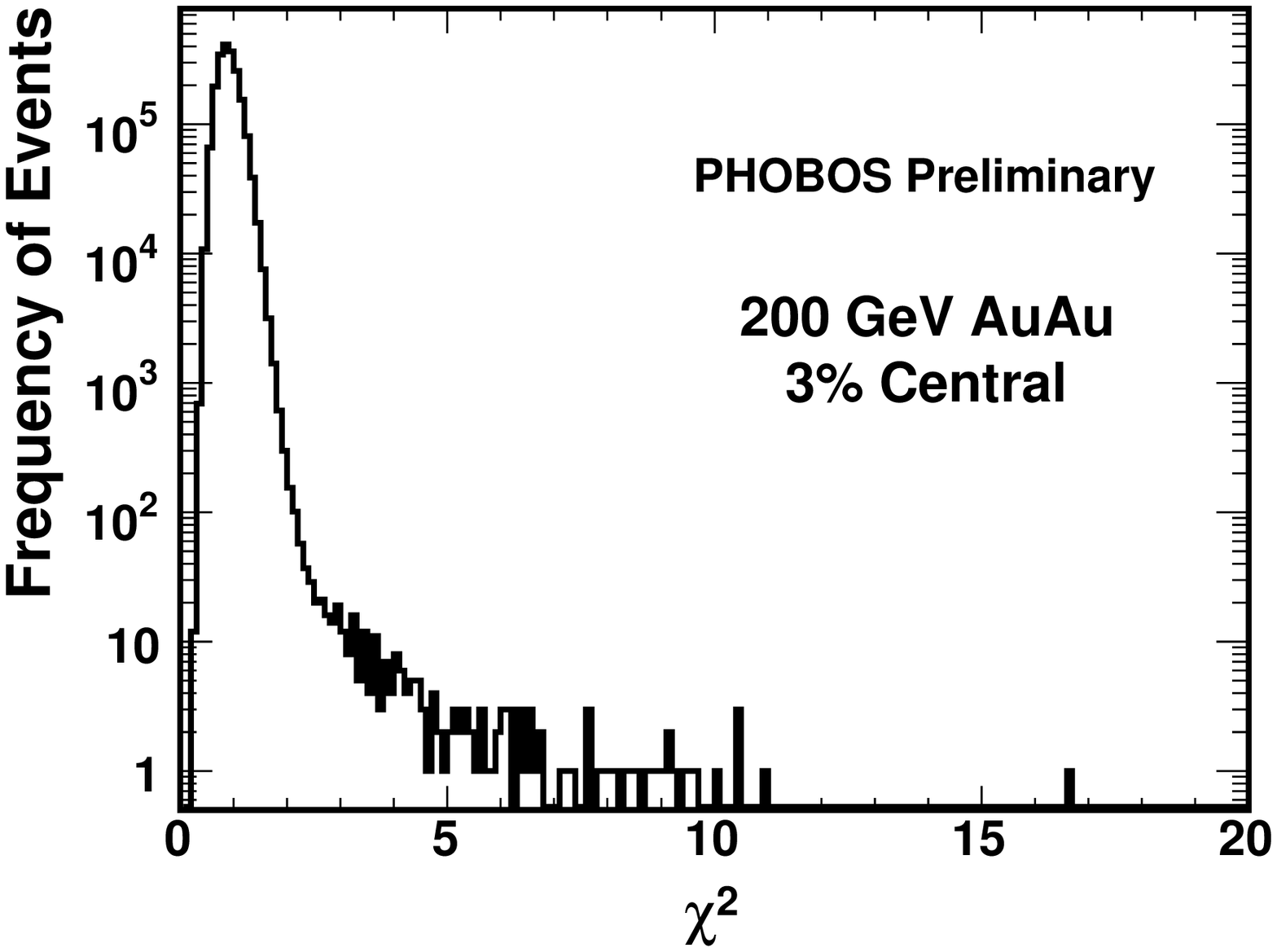}
\caption{Reduced $\chi^2$ distribution for event-by-event comparison of central Au+Au collisions at 200~GeV to ensemble 
average.}
\label{chi2_auau}
\end{minipage}
\end{figure}

The high statistics Au+Au dataset, in combination with the large PHOBOS acceptance, allows us to perform a search for 
events that in a statistically quantifiable way differ from average Au+Au events. Possible mechanisms for the occurrence 
of such events might be the formation of droplets due to supercooling \cite{mishustin} or the formation of disoriented
chiral condensates \cite{wilczek}. While the likelihood of such scenarios is unclear, an unbiased search for ``unusual'' events 
clearly is an important part of the RHIC physics program. The details of our strategy for identifying rare events is 
described in \cite{georgeqm05}. Using the most central two million events of our Au+Au dataset, we determine the 
shape of the average uncorrected $dN/d\eta$ distribution for events in fine bins of vertex position. Similarly, the 
variance around the average shape in bins of $\eta$ is extracted from the data. Using the average shape and variance 
obtained from the data, we then calculate the $\chi^2$ of each individual event relative to the ensemble average. The 
resulting $\chi^2$ distribution is shown in Fig.~\ref{chi2_auau}. The distribution consists of a central core close to 
a $\chi^2/DoF \approx 1$, with a tail out to large $\chi^2$. In total, about 0.01\% of all events are found in a region 
where for purely statistical fluctuations no entries would be expected. Further studies of these unusual events have 
shown that their rate is linearly related to the instantaneous luminosity at which the collisions were recorded. At present, we
therefore have no evidence for the existence of unusual physical fluctuations. Further studies are underway to set quantitative
limits on various physical scenarios.

\subsection{Forward-backward multiplicity correlations}

\begin{figure}[t]
\begin{minipage}[t]{80mm}
\includegraphics[width=7cm]{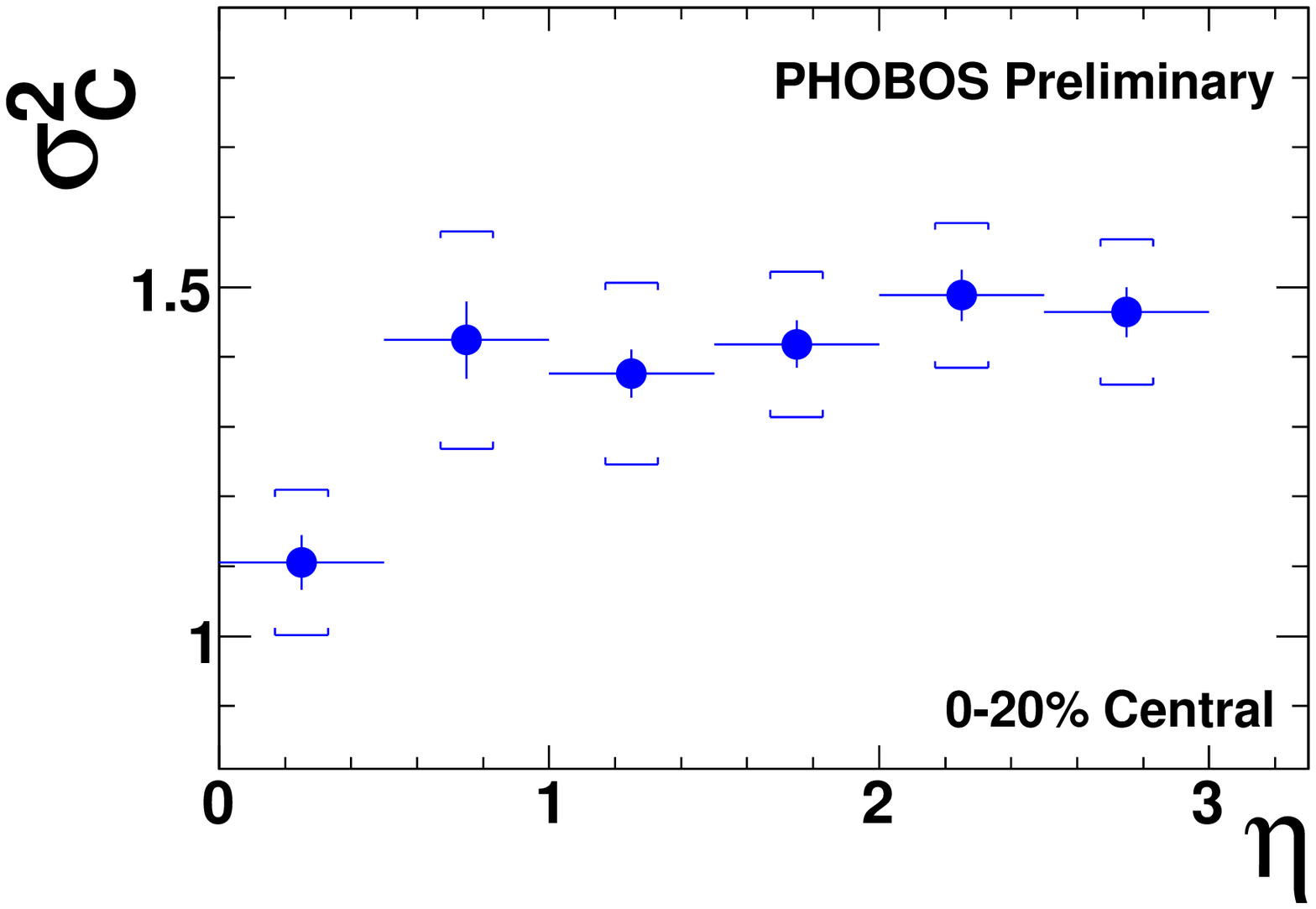}
\caption{Dependence of the fluctuation measure $\sigma_C^2$ on the position $\eta$ for $\Delta \eta = 0.5$ of 
the forward and backward multiplicity bins for central Au+Au collisions at 200 GeV (preliminary). Systematic uncertainties (90\% C.L.) are shown 
as brackets.}
\label{sigmac_eta_auau}
\end{minipage}
\hspace{\fill}
\begin{minipage}[t]{75mm}
\includegraphics[width=7cm]{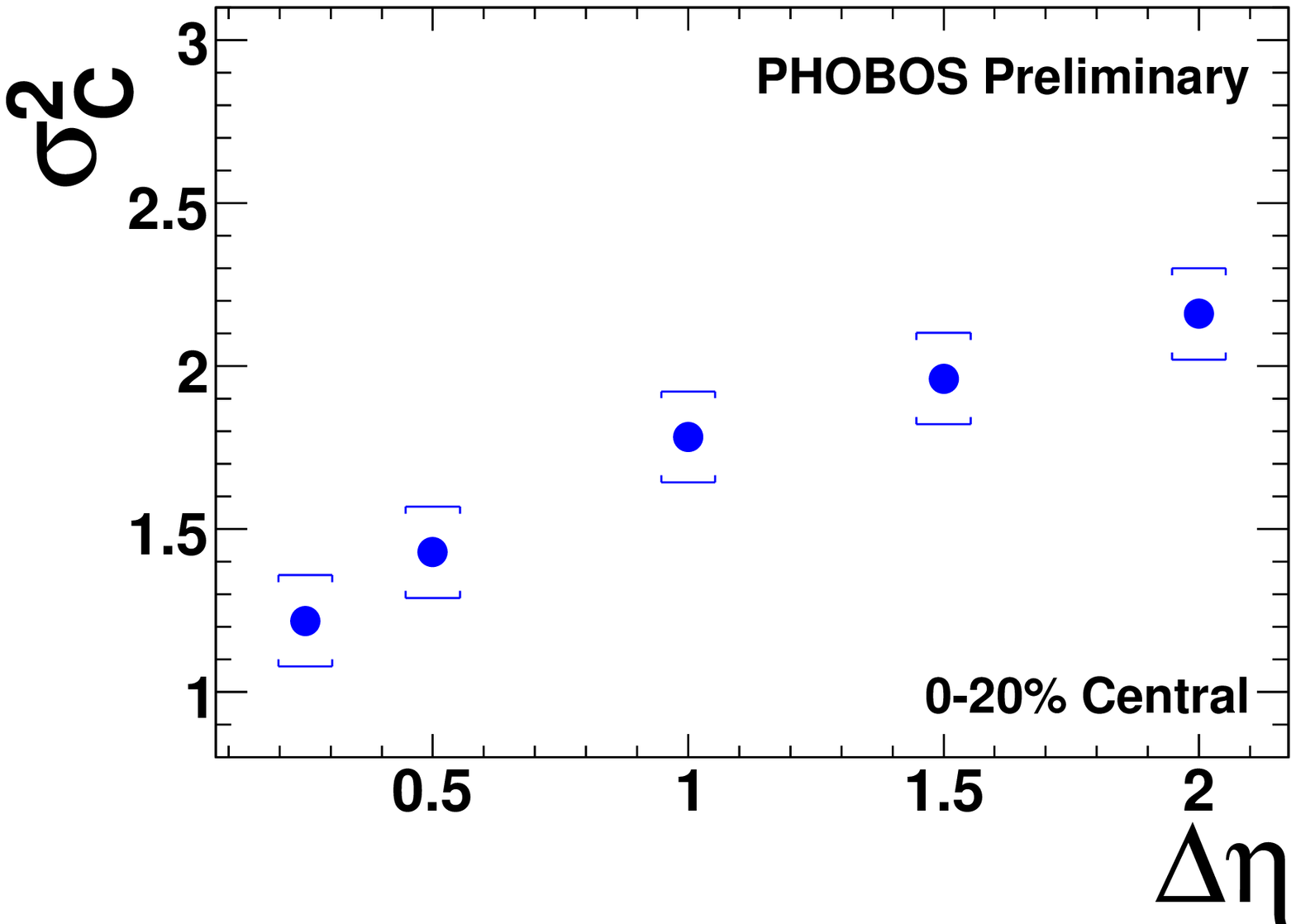}
\caption{Dependence of the fluctuation measure $\sigma_C^2$ on the width $\Delta \eta$ for $\eta = 2.0$ of 
the forward and backward multiplicity bins for central Au+Au collisions at 200 GeV (preliminary). 
Systematic uncertainties (90\% C.L.) are shown as brackets.}
\label{sigmac_delta_eta_auau}
\end{minipage}
\vspace{-5mm}
\end{figure}

Further examination of Fig.~\ref{chi2_auau} shows that the width of the $\chi^2$ distribution around unity is significantly
larger than expected for a purely Poissonian production and detection of charged particles. Thus particles appear to
be produced in a correlated fashion, rather than one-by-one. This can be studied quantitatively using correlations of 
multiplicities in non-overlapping bins of pseudorapidity. We have performed such studies for symmetric pairs of pseudorapidity bins 
centered at values between $0.25 < \eta < 2.75$, varying the bin width from $0.5 < \Delta \eta < 2.0$. To quantify the relative fluctuations 
between the multiplicties $N_F$ in the forward ($\eta > 0$) and $N_B$ in the backward ($\eta < 0$) bins, we define the 
event-wise observable $C = (N_F - N_B)/\sqrt{N_F + N_B}$. This variable has the useful property that its variance 
$\sigma_C^2$ is one for independent particle emission, even when averaged over events from centrality bins of finite widths. 
After correction for detector and acceptance effects, which are described in \cite{zhengwei_mit05,peterqm05}, $\sigma_C^2$ 
can be used to study short range correlations in particle production. If particles are produced as clusters which decay with
a rapidity width smaller than the typical bin width chosen in the analysis, then $\sigma_C^2$ in the absence of other 
correlations will directly correspond to the cluster size $k$, i.e.\ the multiplicity of decay products from each cluster.
In Figs.~\ref{sigmac_eta_auau} and \ref{sigmac_delta_eta_auau}, we show the dependence of $\sigma_C^2$ on the 
position and width of the bins used in our analysis for central Au+Au events. 
The main result is that $\sigma_C^2$ is much larger than unity, in particular for larger $\Delta \eta$, 
indicating that particles in Au+Au collisions are not produced independently, but rather in clusters. The results are reminiscent 
of those obtained in similar analyses for $p+\bar{p}$ collisions \cite{ua5_correlations}. The similarity of the results in A+A and 
$p+\bar{p}$ collisions, as well as the weak energy dependence seen in $p+\bar{p}$, could indicate that the cluster formation
is a phenomenon related to common features at  hadronization. This will be further tested by future studies of multiplicity 
correlations in Au+Au and Cu+Cu collisions as a function of collision energy. For further discussion and a comparison of the present results with event generators, see \cite{peterqm05}.

\section{Summary}

In this paper, we have presented new results from the recent Cu+Cu run at RHIC. The data show that particle production
per participant nucleon is very similar in Cu+Cu and Au+Au collisions, if one selects collisions with equivalent number
of participants. This is true not only for the mid-rapidity particle density, but also for $p_T$ spectra and the 
shape of the pseudorapidity distributions. Like Au+Au collisions at RHIC, the Cu+Cu data also exhibit several 
intriguing scaling relationships, including the extended longitudinal scaling of pseudorapidity distributions and 
elliptic flow, and the factorization of energy and centrality dependence. The first data on elliptic flow in Cu+Cu reveal
a shape of $v_2(\eta)$ that is similar to observations in Au+Au, with a surprisingly large magnitude of $v_2$ near mid-rapidity
relative to the expected average initial state anisotropy. We have argued that the result can be quantitavily understood,
when taking fluctuations in the initial state transverse geometry into account. This leads to a universal scaling of 
$v_2$ relative to the properly defined ``participant eccentricity'' in Cu+Cu and Au+Au data over a large range in 
collision energy.
We also presented examples of the ongoing systematic study of particle production at very low transverse momenta. 
For Au+Au, the low $p_T$ data agree well with extrapolations from higher $p_T$ assuming a radially expanding source.
We expect further insight into the underlying mechanisms of particle production from measurements of event-by-event fluctuations. 
Results were shown that revealed significant, cluster-like,  correlations in final state multiplicity production, 
but no indication for the presence of unusual, large scale fluctuations. Overall, our data point to the importance of understanding
the dynamics of the very early stage of the collision, where the scaling features observed in the final state hadron production
and anisoptropies are established.

This work was partially supported by U.S. DOE grants
DE-AC02-98CH10886,
DE-FG02-93ER40802,
DE-FC02-94ER40818,  
DE-FG02-94ER40865,
DE-FG02-99ER41099, and
W-31-109-ENG-38, by U.S.
NSF grants 9603486, 
0072204,            
and 0245011,        
by Polish KBN grant 1-P03B-062-27(2004-2007),
by NSC of Taiwan Contract NSC 89-2112-M-008-024, and
by Hungarian OTKA grant (F 049823).

\end{document}